\documentclass[DIV=12]{scrartcl}
\usepackage[a4paper, margin=2cm]{geometry}
\usepackage{graphicx} % Required for inserting images
\usepackage{hyperref}

\usepackage[backend=biber,
style=apa,
citestyle=authoryear-comp,
natbib=true,uniquename=false,
sorting=nyt]{biblatex}
\title{Rewiring of adult neural networks -- learning from structural plasticity across multiple scales}

\input{preamble_author_affilition}

\author*{Michael Fauth}
\affiliation{Third Institute for Physics, Georg August University Göttingen, Göttingen, Germany}

\author*{Han Lu}
\affiliation{Simulation \& Data Lab Neuroscience, Jülich Supercomputing Centre (JSC), Forschungszentrum Jülich GmbH, Jülich, Germany}

\author*{Jannik Luboeinski}
\affiliation{Simulation \& Data Lab Neuroscience, Jülich Supercomputing Centre (JSC), Forschungszentrum Jülich GmbH, Jülich, Germany}
\affiliation{Campus Institute Data Science (CIDAS), University of Göttingen, Göttingen, Germany}

\author*{Johanna Senk}
\affiliation{Sussex AI, School of Engineering and Informatics, University of Sussex, Brighton, United Kingdom}
\affiliation{Institute for Advanced Simulation (IAS-6), Jülich Research Centre, Jülich, Germany}

\author{Simone Albani}
\affiliation{Institute of Neuroscience  and Medicine (INM-9), Jülich Research Center, Germany}

\author{Valentina Armuzza}
\affiliation{Simulation \& Data Lab Neuroscience, Jülich Supercomputing Centre (JSC), Forschungszentrum Jülich GmbH, Jülich, Germany}

\author{Paolo Carloni}
\affiliation{Institute of Neuroscience  and Medicine (INM-9), Jülich Research Center, Germany}

\author{Richmond L. Crisostomo}
\affiliation{Laboratory for Biomicrotechnology, Department of Microsystems Engineering–Institut für Mikrosystemtechnik, Faculty of Engineering, University of Freiburg, Freiburg, Germany}
\affiliation{Bernstein Center Freiburg, University of Freiburg, Freiburg, Germany}

\author{Hermann Cuntz}
\affiliation{Ernst Strüngmann Institute (ESI) for Neuroscience of the Max Planck Society, Frankfurt am Main, Germany
Computer-Based Modelling in the field of 3R Animal Protection, Faculty of
Medicine, Justus Liebig University Giessen, Giessen, Germany}

\author{Ulrich Egert}
\affiliation{Laboratory for Biomicrotechnology, Department of Microsystems Engineering–Institut für Mikrosystemtechnik, Faculty of Engineering, University of Freiburg, Freiburg, Germany}
\affiliation{Bernstein Center Freiburg, University of Freiburg, Freiburg, Germany}

\author{Eugenio	F. Fornasiero}
\affiliation{Department of Neuro- and Sensory Physiology, University Medical Center Göttingen, Göttingen, Germany}
\affiliation{Department of Life Sciences, University of Trieste, Trieste, Italy}

\author{Tobias Gemmeke}
\affiliation{Integrated Digital Systems and Circuit Design (IDS), RWTH Aachen University, Aachen, Germany}

\author{Claus C. Hilgetag}
\affiliation{Institute of Computational Neuroscience, Hamburg Center of Neuroscience, University Medical Center Hamburg-Eppendorf (UKE), Hamburg University, Hamburg, Germany}

\author{Sabine M. Hölter}
\affiliation{Institute of Experimental Genetics, Helmholtz Zentrum München, German Research Center for Environmental Health, Neuherberg, Germany}

\author{Patrick Inoue}
\affiliation{Department of Computer Science, Chemnitz University of Technology, Chemnitz, Germany}

\author{Björn Kampa}
\affiliation{Department of Systems Neurophysiology, Institute for Biology II, RWTH Aachen University, Aachen, Germany}
\affiliation{JARA-BRAIN Institute INM-10, Forschungszentrum Jülich, Germany}

\author{James C. Knight}
\affiliation{Sussex AI, School of Engineering and Informatics, University of Sussex, Brighton, United Kingdom}

\author{Susanne Kunkel}
\affiliation{Neuromorphic Software Ecosystems (PGI-15), Jülich Research Centre, Jülich, Germany}

\author{Thomas Nowotny}
\affiliation{Sussex AI, School of Engineering and Informatics, University of Sussex, Brighton, United Kingdom}

\author{Samora Okujeni}
\affiliation{Laboratory for Biomicrotechnology, Department of Microsystems Engineering–Institut für Mikrosystemtechnik, Faculty of Engineering, University of Freiburg, Freiburg, Germany}
\affiliation{Bernstein Center Freiburg, University of Freiburg, Freiburg, Germany}

\author{Nicola Palomero-Gallagher}
\affiliation{Institute of Neuroscience and Medicine (INM-1), Research Centre Jülich, Jülich, Germany}
\affiliation{C. \& O. Vogt Institute for Brain Research, Heinrich-Heine-University, Düsseldorf, Germany}

\author{Mariia Popova}
\affiliation{Institute of Computational Neuroscience, Hamburg Center of Neuroscience, University Medical Center Hamburg-Eppendorf (UKE), Hamburg University, Hamburg, Germany}

\author{Petra Ritter}
\affiliation{Berlin Institute of Health at Charité — Universitätsmedizin Berlin, Charité, Berlin, Germany}

\author{Giulia Rossetti}
\affiliation{Institute of Neuroscience  and Medicine (INM-9), Jülich Research Center, Germany}
\affiliation{Department of Neurology, University Hospital Aachen, RWTH Aachen University, Aachen, Germany}

\author{Michael Schirner}
\affiliation{Berlin Institute of Health at Charité — Universitätsmedizin Berlin, Charité, Berlin, Germany}

\author{Marc Spehr}
\affiliation{Department of Chemosensation, Institute for Biology II, RWTH Aachen University, Aachen, Germany}

\author{Leon Stefanovski}
\affiliation{Berlin Institute of Health at Charité — Universitätsmedizin Berlin, Charité, Berlin, Germany}

\author{Gaia Tavosanis}
\affiliation{Chair of Developmental Biology/Bio2, RWTH Aachen University, Aachen, Germany}

\author{Agata Witkowska}
\affiliation{Leibniz-Forschungsinstitut für Molekulare Pharmakologie (FMP), Berlin, Germany}
\affiliation{Neurocure Cluster of Excellence, Charite-Universitätsmedizin Berlin, Berlin, Germany}

\author{Felix Wolf}
\affiliation{Department of Computer Science, Technical University of Darmstadt, Darmstadt}

\author{Willem Wybo}
\affiliation{Peter Grünberg Institute 15 – Neuromorphic Software Ecosystems (PGI-15), Jülich Research Center, Aachen, Germany}

\author**{Sandra Diaz-Pier}
\affiliation{Simulation \& Data Lab Neuroscience, Jülich Supercomputing Centre (JSC), Forschungszentrum Jülich GmbH, Jülich, Germany}

\author**{Maximilian Lenz}
\affiliation{Hannover Medical School, Institute of Neuroanatomy and Cell Biology, Hannover, Germany}
\affiliation{Center for Systems Neuroscience, Hannover, Germany}

\author**{Abigail Morrison}
\affiliation{Institute for Advanced Simulation (IAS-6), Jülich Research Centre, Jülich, Germany}
\affiliation{Software Engineering, Faculty of Computer Science, RWTH Aachen University, Aachen, Germany}

\author**{Florian Röhrbein}
\affiliation{Department of Computer Science, Chemnitz University of Technology, Chemnitz, Germany}

\author**{Stefan Rotter}
\affiliation{Bernstein Center Freiburg, University of Freiburg, Freiburg, Germany}
\affiliation{Faculty of Biology, University of Freiburg, Freiburg, Germany}

\author**{Johannes Schemmel}
\affiliation{Institute of Computer Engineering, Heidelberg University, Heidelberg, Germany}

\author**{Christian Tetzlaff}
\affiliation{Department of Neuro- and Sensory Physiology, University Medical Center Göttingen, Göttingen, Germany}

\author**{Andreas Vlachos}
\affiliation{Department of Neuroanatomy, Institute of Anatomy and Cell Biology, Faculty of Medicine, University of Freiburg, Freiburg, Germany}

\date{} % to not show today's date

\begin{document}

\maketitle

\newpage
{\small \tableofcontents}
\newpage

\addsec{Motivation and overview}

\begin{figure}[bh!]
    \centering
    \includegraphics[width=0.8\linewidth]{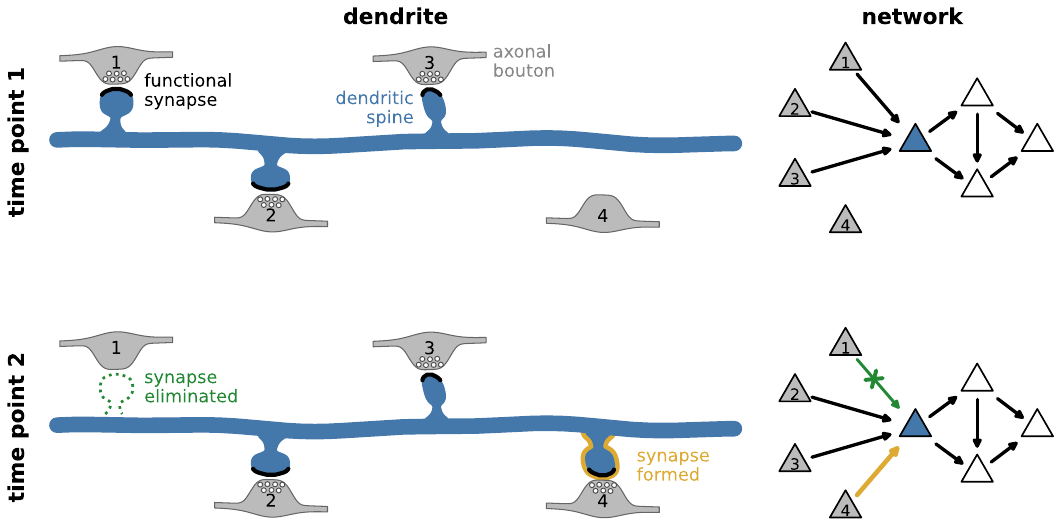}
    \caption{
    \textbf{Illustration of structural plasticity:}
    Functional chemical synapses linking dendritic spines with axonal boutons are continuously formed and eliminated.
    The synapse modifications locally at a neuron's dendrite lead to a change in connectivity at the network level.
    }
    \label{fig:structplast}
\end{figure}
Neurons in biological and artificial systems form complex networks, whose structure determines information processing. In the brain, the connections within these networks are realized predominantly by chemical synapses. Strikingly, these connections are not static but highly dynamic: On the one hand, synapses themselves change over time and adapt their transmission efficacy as well as their morphology via synaptic plasticity, for example long-term potentiation (LTP), long-term depression (LTD) or spike-timing dependent plasticity (STDP)~\citep{Bliss1973, Matsuzaki2004,Zhou2004,Dudek1992,Bi1998}. On the other hand, \textbf{structural plasticity} is the process of creating new synapses or entirely removing old ones, thus making the network topology dynamic (see Fig.~\ref{fig:structplast} for a sketch). The literature also refers to this phenomenon as synaptic turnover, synaptogenesis, rewiring or architectural plasticity -- with partially different connotation. Structural plasticity is not to be confused with structural synaptic plasticity, which refers to the morphological remodelling of individual synapses, which is also related to rewiring as discussed later.

Structural plasticity is not only essential for the development of neural networks during ontogenesis~\citep{Huttenlocher1982,Huttenlocher1984}, but also found to occur in adult neural networks~\citep{Trachtenberg2002,Chen2012,VanVersendal2012} and to be experience-dependent~\citep{Xu2009, Yang2009, Moczulska2013} or help the recovery from lesions~\citep{Cheetham2008,Keck2008}, which highlights it as an important mechanism the brain can employ to learn and adapt to its environment.
Yet, a clear understanding how and to what extent adult structural plasticity adapts neural information processing is missing. 

A central challenge is that structural plasticity is inherently multiscale: molecular and subcellular remodeling determines the formation, stabilization and elimination of individual connections, while the functional consequences of these processes emerge at the level of circuits and networks. Neuronal activity typically evolves on the order of milliseconds, but structural plasticity requires hours or even days and beyond to unfold. In experimental neuroscience, substantial progress has been made in constructing large-scale connectomes~\citep{Dorkenwald2024, Shapson2024, Microns2025}, but they are recorded at a single time point and thus only provide a static snapshot of the dynamic network topology. Hence, temporal dynamics of macroscopic connectomes can only be extrapolated from multiple samples collected at discrete time points. On the other hand, longitudinal imaging studies permit tracking of the dynamics of synapses over hours to days, but this approach only accommodates parts of the networks (e.g., individual postsynaptic neurons in two-photon-imaging~\citep[][, compare dendritic spines dynamics similar to Fig.~\ref{fig:structplast} left]{Trachtenberg2002} or only the locations of synapses without information about the pre- and postsynaptic neurons when labeling synaptic structures~\citep{Verstraelen2020}).
Due to this multiscale property across both space and time, comprehending structural plasticity is a hard problem.
Mathematical and computational models provide a means to bridge these spatial and temporal scales by extracting principles from observations at the molecular, synaptic or cellular level and testing their emergent consequences in networks. Yet, so far structural plasticity has received limited attention from modelers due to its ambiguous definition given the high dimensionality of morphological data, and the view that structural plasticity is mathematically already covered by synaptic plasticity as removing a synapse or having a synapse with no transmission efficacy (weight zero) creates a functionally equivalent network in first-order approximation. This functional equivalence at one point in time does however not imply that this equivalence holds over time, and it certainly does not imply biological equivalence. Along this line, structural plasticity of the network usually occurs on much slower timescales than synaptic plasticity (see, e.g.,~\cite{Kwon2011} vs.~\cite{Bi1998}) and may therefore serve different functions. Moreover, in biology as well as in artificial networks and neuromorphic chips, synapses are a limited resource that is energetically costly~\citep{Harris2012,Loihi2TechBrief,Scholze2026,Knight2026} such that not every connection can be maintained. Hence, structural plasticity rather complements changes in synaptic strength and thereby allows the matured brain to adapt to different computational and resource requirements. Along this line, understanding network self-organization by structural plasticity also has multiple translational perspectives, which have emerged in the recent years: on the one hand, it could be translated to artificial neuronal networks, for example to improve network architectures~\citep{Lecun1989,Achterberg2023} and allow for better energy efficiency in artificial intelligence and neuromorphic computing~\citep{Bellec2018, Frankle2018, Petschenig2023, Tmamna2024}. Moreover, understanding how the brain adapts its connectivity is required for building large-scale network models that can provide a holistic understanding of the brain~\citep{DiazPier2016,Einevoll2019,Amunts2024,Plesser2025}. 
Besides this, understanding network adaptation will also allow the development of interventions and therapies in psychology and neurological disease~\citep{forrest2018dendritic}. 
To this end, knowledge about therapy effects on molecular and cellular scale needs to be combined to predict network and brain scale effects.

\begin{figure}
    \centering
    \includegraphics[width=\linewidth]{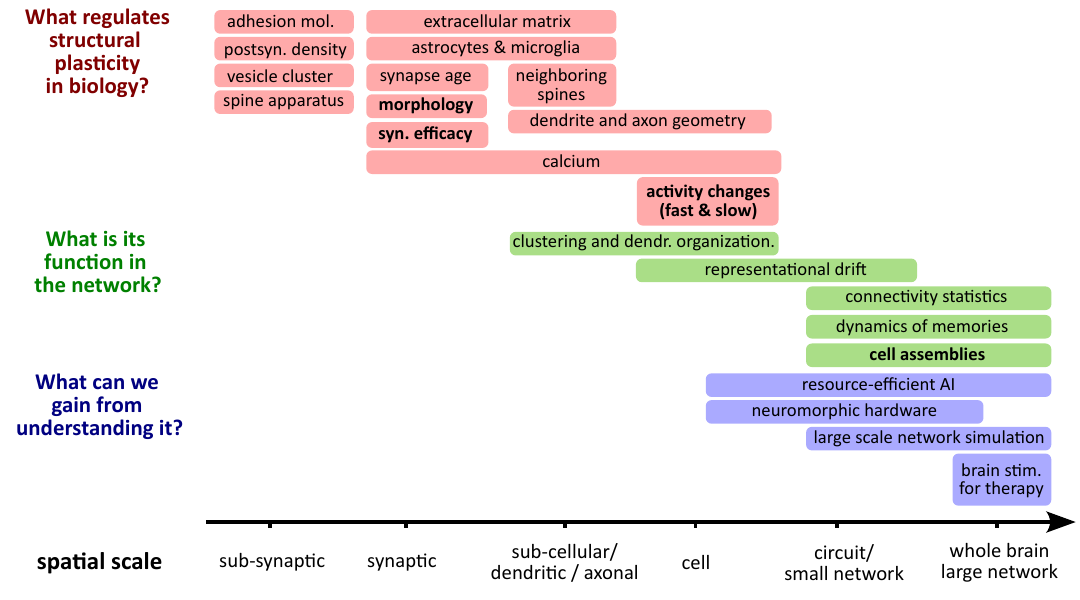}
    \caption{Overview of the discussed questions and research topics of structural plasticity.}
    \label{fig:overview}
\end{figure}

Structural plasticity appears to be one of a few topics in neuroscience research that are at the same time integral and conjunctive for the field, but still understudied.
Recent progress in experimental techniques, related theoretical work, and the development of simulation technology indicates that the near future allows us to formulate and address complex research questions with multidisciplinary approaches that were out of reach a decade ago.
Taken together, structural plasticity is a timely and relevant topic, and we hope to see progress in answering its three major research questions in the time to come (Fig.~\ref{fig:overview}):
\begin{enumerate}
    \item What regulates structural plasticity in biological neural networks?
    \item What are the functional consequences and advantages of structural plasticity in adult neural networks?
    \item How can principles of structural plasticity be translated into clinical and technological applications?
\end{enumerate}
In the following we will review the existing and missing work on these questions and argue that answering them requires a collective effort of experimental, computational and translational researchers.

\newpage
\section{What regulates structural plasticity in biological networks?}
\label{sec:regulation}

\subsubsection{State of the art}
As a first challenge, we need to identify the underlying principles that govern network reconfiguration in mature circuits. In the following, we will hereby focus mainly on excitatory synapses which typically reside on dendritic spines~\citep{Yuste2010} - small mushroom shaped protrusions with a neck and a head -- which can be easily identified through two-photon imaging when labeling the whole cell. From such imaging studies we know that, even in mature circuits, synapses are continuously formed and eliminated~\citep{Trachtenberg2002}. Various lines of research have identified factors that help to predict whether synapse formation or elimination are more or less likely at certain locations or time points, which we will summarize in the following. However, so far, it is not possible to predict where and when exactly synapses are formed, stabilized or eliminated. Hence, synapse formation and elimination currently need to be treated as stochastic processes, while factors discussed below modulate the probabilities or rates of these processes.

\paragraph{Long-lasting alterations of network activity}
While chemical synapses undergo continuous turnover in the brain under baseline conditions, one of the most important factors regulating structural plasticity is neural activity (see~\cite{Butz2009, Fauth2016} for reviews).
As the formation and removal of synapses in turn influences neural activity, such an activity-dependent structural plasticity leads to a feedback loop between connectivity and activity, which can be either negative or positive. In a homeostatic, negative-feedback framework, an increase in activity would lead to a decrease in connectivity and thereby drive activity back towards its original state. Likewise, decreased activity would lead to increased connectivity, which in turn restores activity. Structural plasticity implementing such a negative feedback is thus called \textbf{homeostatic structural plasticity} as it helps to achieve activity homeostasis (see~\citet{Butz2009} for a review).

Evidence for this form of structural plasticity comes from chronic alterations of network activities by various drugs. For example, blocking neural activity for more than 8 hours leads to the growth of new dendritic spines~\citep{Wolff1978, Dalva1994, Kirov19999} and increasing neural activity levels leads to a loss of dendritic spines~\citep{Müller1993, Drakew1996, Seeburg2008, Goold2010, Fiore2014,, Chowdhury2017, Mendez2018, Mendez2018, Moulin2019, Kesaf2020}. Notably, already blocking calcium influx through NMDA-channels stabilizes spines~\citep{Bock1999} and increases spine number~\citep{Yu2013, Chen2015} while a direct application of NMDA leads to spine loss~\citep{Halpain1998}. Together this suggest that either calcium is an important messenger that regulates activity-homeostasis through structural plasticity, or structural plasticity serves the homeostatic regulation of calcium levels~\citep{Kater1989, Segal2000} and activity homeostasis is a result of this.

In contrast to the above findings, there is also a substantial body of studies showing that chronic inhibition of neural activity triggers non-homeostatic spine loss~\citep{Tyler2003, Wallace2004, Wierenga2005, Thiagarajan2005}, indicating that homeostatic plasticity only operates within certain bounds or exhibits non-linear activity-dependence.

On the other hand, homeostatic structural plasticity can also be observed in situations where adult networks face drastic changes in their activity levels. For example, when neurons lose their input -- such as after denervation~\citep{Vlachos2012, Platschek2016}, monocular deprivation~\citep{Hofer2009}, whisker trimming~\citep{Vees1998, Chen2015}, or after retinal lesions~\citep{Keck2008} -- they increase the number of spines in response in order to compensate input loss. High activity levels, as for example during epileptic seizures, in turn decrease the number of spines~\citep{Scheibel1974, Müller1993, Isokawa1998}.

Taken together, long-lasting perturbations of network activity can engage homeostatic structural remodeling that may contribute to restoring activity toward a target or physiological range. Hereby, the direction of the structural changes may exhibit a non-linear or even non-monotonic dependence on long-lasting activity alterations.

\paragraph{Fast changes in activity}
In comparison to the impact of long-term activity alterations, fast and transient activity changes can also induce structural remodeling, particularly in the context of stimuli that induce synaptic plasticity such as LTP or LTD (see \cite{Fauth2016} for a review). 
For example, an LTP-inducing stimulus or simply the release of glutamate can increase the number of spine precursors (filopodia) and synapses already a few minutes after stimulation~\citep{Lee1980, Chang1984, Moser1994, Trommald1996, Engert1999, Kwon2011}.  Furthermore, LTP-inducing stimulation aids the stabilization and maturation of nascent dendritic spines~\citep{Hill2013}. At the same time, axonal boutons are formed around a few minutes after LTP induction~\citep{Nikonenko2003, Ninan2006}, indicating the formation of new synapses~\citep{Watson2015}. Similar to LTP, this implements the ``fire-together-wire-together'' principle proposed by Donald Hebb~\citep{Hebb1949, Lowel1992, Shatz1992}, and this mechanism is accordingly termed \textbf{Hebbian structural plasticity}. In contrast to the homeostatic variant, this creates a positive feedback between activity and connectivity, potentially leading to runaway dynamics: when cell A takes part in triggering an action potential in cell B, Hebbian structural plasticity will lead to the formation of more synapses such that the influence of cell A on cell B dynamics becomes stronger. It is therefore likely that Hebbian mechanisms act together with  homeostatic structural plasticity or resource constraints that impose growth boundaries to ensure network stability.

On the other hand, LTD-inducing stimulation leads to a loss of dendritic spines~\citep{Naegerl2004,Yu2013} and increases axonal dynamics~\citep{Paola2006}, and, thus, synapse loss~\citep{Becker2008}. 
Thus, both LTP and LTD-inducing stimuli can trigger structural changes whose direction parallels the respective changes in synaptic efficacy.
Taken together with the observations in response to long-term activity alterations discussed above, we conclude that structural plasticity exhibits complex behaviour across temporal scales with positive feedback turning into negative feedback and further investigations how these interact are needed.

\paragraph{Spine size and synaptic weight}
As the above observations already suggest,  synaptic and structural plasticity are linked~\citep{Segal2005}: The stability of synapses has been found to depend on head size of the corresponding dendritic spine -- that is, synapses on larger spines are more stable~\citep{Grutzendler2002, Majewska2006, Yasumatsu2008, Loewenstein2015, Pfeiffer2018}. The spine head size, in turn, is correlated with the size of the postsynaptic density~\citep{Borczyk2019, Sun2021, Wegner2022} and, thus, ultimately with the synaptic strength (or weight) as measured by the postsynaptic currents~\citep{Matsuzaki2001, Knott2006, Zito2009}. Accordingly, synapses that transmit more current are more stable~\citep{Holtmaat2005, LeBe2006, Rais2026}.
During LTP and LTD, synapses increase and decrease their weight and dendritic spines undergo expansion and shrinkage, respectively~\citep[structural LTP and LTD;][]{Matsuzaki2004, Zhou2004}, which in turn implies that they become more or less stable. Thus synaptic weight (or spine size) and synapse stability are tightly coupled giving rise to a \emph{weight- or size-dependent structural plasticity} that may also explain parts of the above discussed Hebbian effects. 

\paragraph{Spine morphology and actin dynamics} 
Beyond the spine head size, multiple further morphological properties of the spine influence synapse stability. Already two-photon data shows that geometric properties such as the ellipticity of the spine head and its distance from the dendrite have a small but significant influence on the stability~\citep{Loewenstein2015}. Super-resolution techniques further revealed that also smaller widths and lengths of the spine neck correlate with increased spine elimination~\citep{Steffens2021}. 

These morphological features as well as the spine head size are not only heavily altered during synaptic plasticity~\citep{Matsuzaki2004, Bosch2014, Sun2021, Tonnesen2014}, but also undergo massive spontaneous changes~\citep{Fischer1998, Yasumatsu2008, Steffens2021}. Hence, morphology can only be partly controlled through network activity~\citep{Dvorkin2016}, which indicates that parts of the randomness inherent to structural plasticity and synaptic turnover may emerge from such spontaneous morphology changes.
Thus, a deeper insight into the random and non-random morphology dynamics may be key to understanding the nature of structural plasticity.

A major determinant of spine morphology is the actin cytoskeleton. It stabilizes the spine neck by forming stable actin-spectrin rings~\citep{Baer2016} and determines spine head geometry and its (random) dynamics by exerting expansive forces~\citep{Fischer1998}. To this end, actin interacts with multiple actin binding proteins to form filament networks~\citep{Korobova2010} or crosslinked gels~\citep{Honkura2008, Okamoto2007}. Blocking actin polymerization stops rapid morphology changes~\citep{Fischer1998}, indicating that actin polymerization contributes to spontaneous (activity-independent) shape fluctuations. Interference with filament polymerization or the actin binding proteins consequently also influences synapse stability (e.g., ARP2/3:~\cite{Wegner2008}; ADF/Cofilin:~\cite{Rust2015}; NRB1:~\cite{Zito2004}).
On the other hand, actin binding proteins are modulated  by neural activity~\citep{Matsuzaki2004, Zhou2004, Bosch2014} giving rise to the above described activity-dependent morphology and spine size changes~\citep{Nakahata2018,BonillaQuintana2020, BonillaQuintana2021,Thomas2025, BonillaQuintana2024}.

Moreover, energetic cost of membrane deformation, for example during spine initiation and retraction, are determined by properties of the plasma membrane itself, such as material properties, curvature, and lipid composition, that, in turn, depend on the membrane's lipid and protein composition \citep{Haucke2018,Alimohamadi2021}. 

Finally, membrane-associated proteins of the BAR domain superfamily, that act as mechanochemical hubs coupling membrane state and curvature generation to actin remodeling, furthermore establish a feedback between membrane shape and state and intracellular signalling~\citep{Kessels2015, Khanal2021, Kessels2021}.

Taken together, spine stability is closely linked to spine morphology, which undergoes both spontaneous and activity-dependent remodeling mediated by  actin  dynamics, hence establishing a direct link between self-organization on the network scale and molecular processes at the individual synapse.

\newpage
\paragraph{Age}
Additionally, two-photon studies indicate that the age of the spines -- that is, how long it has persisted in previous imaging sessions -- positively influences spine stability~\citep{Loewenstein2015}. Similarly preexisting axonal boutons are more stable than newly formed ones~\citep{Qiao2016}. Thus, there may be either an ongoing maturation process that makes synapses more and more stable or synapses have further unknown properties determining their stability such that synapses that survive longer are also more likely to be more stable (survivors bias). Anyway, the age-dependence of synapse stability supports the notion that there may be a stable backbone of synapses that support network function~\citep{Kasai2003, Chambers2017, Mongillo2017}.

However, it is unlikely that spines have a built-in age-tracking mechanism, but rather that structural spine plasticity is regulated by one or multiple molecular or morphological correlates of spine age, which may be amongst the factors discussed here. One such molecular correlate may be the age of the synaptic proteins themselves: protein lifetimes in the brain differ widely across proteins and subcellular compartments, including a set of extremely long-lived proteins~\citep{Fornasiero2018}, and synaptic protein turnover correlates with synaptic morphology, including pre- and postsynaptic size~\citep{Li2024}.

\paragraph{Synaptic clustering and diffusion between neighboring spines}
Besides the morphology and age of the spine itself, also the local neighborhood plays an important role -- that is, spines exhibit clustered (structural) plasticity. For example, spines gained and stabilized upon learning or LTP~\citep{Fu2012,Reiner2015, Frank2018, Hedrick2024}  or the consecutive consolidation during sleep~\citep{Yang2014} tend to cluster within a few micrometers along the dendrite, or at least at the same dendritic branch. Multiple causes for this have been proposed (reviewed in~\cite{Kastellakis2019} and~\cite{Debello2017}). One possibility is that the morphology and organization of neurons entails that close-by synapses have correlated presynaptic activity (e.g., inputs are from neighboring neurons of the same topological map or even from same presynaptic axon) and are stabilized by the above discussed plasticity mechanisms. This may be  aided by local dendritic plateau potentials. Another reason would be that close-by synapses share stabilizing signals, for example through diffusion within the dendrite. Along this line, regulators of the actin dynamics are known to diffuse to neighboring spines~\citep{Nishiyama2015, Chater2021, Oh2015} and the local propagation of calcium signals influences plasticity of neighboring spines~\citep{Lee2016, Oh2015,Shafiee2026}. Moreover, the synthesis of plasticity related proteins, which is necessary for long-lasting synaptic plasticity, can be realized at a local level~\citep{Chirillo2019,Rangaraju2017,  Sun2021b}. This also influences spine stabilization~\citep{Sun2021b}, likely via the above discussed weight- and size-dependence.  Similarly, also the elimination of spines depends on diffusible signals~\citep{Yasuda2021, NagappanChettiar2024}. 

Hence, structural plasticity is not purely synapse-autonomous but is \emph{strongly influenced by the local dendritic neighborhood}. The underlying molecular mechanisms and the functional consequences in the network still need clarification (but see Section~\nameref{sec:dendrites}).

\paragraph{Local supply structures}
Another reason for clustered structural plasticity may be the heterogeneous distribution of cell internal structures. On the one hand, there is the smooth endoplasmic reticulum (SER), which is involved in intracellular calcium storage and signaling and forms a specialized  spine apparatus in a subset of dendritic spines. The presence of SER in the dendrite corresponds to locally increased spine densities~\citep{Spacek1997, Chirillo2019} and spines containing spine-aparatus-marker synaptopodin are more stable than the ones without~\citep{Yap2020}. Moreover, synaptopodin is implicated in the regulation of homeostatic rewiring~\citep{Vlachos2013}.
On the other hand, there are mitochondria that supply the cell with energy. Also here, axonal boutons containing mitochondria are more stable~\citep{Lees2020} and spines close to mitochondria are larger~\citep{Thomas2023}, and hence more stable. 

Hence, heterogeneous distribution of intracellular organelles such as SER/spine apparatus or mitochondria, may give rise to a \emph{spatial heterogeneity in structural plasticity}, potentially contributing to the synaptic clustering discussed above. Strikingly, both organelles are dynamic and can be influenced by neural activity~\citep{Lin2015,Ng2014, Perez2020}, such that networks may be able to self-organize the heterogeneity of structural plasticity.

\paragraph{Neuron morphology}
Further spatial heterogeneity is introduced by the morphology of the neurons, in particular their axons and dendrites. A synapse between two neurons can only be formed if their axonal and dendritic arbors come into sufficiently close apposition to permit synapse formation. Such apposition sites have been termed potential synapses \citep{Chklovskii2004, Stepanyants2002, Stepanyants2005}.  The spatial distribution of these potential synapses constrains activity-dependent structural plasticity such that a heterogeneous distribution of the potential synapses between two neurons over the dendritic arbour could result in heterogenous structural changes or even clustered spine formation.

Under baseline conditions, axons and dendrites do not remodel strongly~\citep{Trachtenberg2002, Grutzendler2002, Paola2006, Stettler2006}. However, following major perturbations, axonal and dendritic remodeling can alter the space of potential connections~\citep{Butz2009} and their respective function~\citep{Platschek2016,Cuntz2021}.
This may constitute a form of structural metaplasticity by changing the capacity for subsequent structural plasticity.

\paragraph{Cell adhesion and synaptic structures}

Synapse formation is thought to be initiated by synaptic organizing molecules, most prominently cell adhesion molecules (CAMs, see~\cite{Suedhof2021} for a review). For example, neurexins and neuroligins -- CAMs  that bind each other and are found in the pre- and postsynaptic membrane, respectively -- can already on their own induce pre- and postsynaptic specialisations in heterologous synaptogenesis assays~\citep{Scheiffele2000, Nam2005}. At a newly forming synapse, they are assumed to make contact with each other, which hinders them from diffusing away. Inside the pre- and postsynapse, they then organize scaffolding complexes and multi-protein signaling which promotes pre- and postsynaptic differentiation~\citep{Reshetniak2025}. This ultimately leads to the formation of a vesicle cluster that implements neurotransmitter release, and the postsynaptic density (PSD) that hosts the respective receptors to convert the chemical signals into postsynaptic currents. In line with the relation of PSD size, spine size and spine stability discussed above, these structures are not only functionally relevant, but provide a structural scaffold that stabilizes the synapse. For example, constituents of the PSD contribute to the activity-dependent stabilization of dendritic spines (e.g., PSD95:~\cite{Ehrlich2007, Cane2014, Taft2014}; although spine loss does not require prior PSD loss, see~\cite{Woods2011}). Conversely, overexpression of individual postsynaptic scaffolding proteins can be sufficient to induce spines (e.g. SHANK:~\cite{Roussignol2005}), while overexpression of components of the presynaptic vesicle cluster increases spine stability~\citep{Bosch2024}. 
Beyond these individual proteins, multiplexed super-resolution imaging revealed that synapses are highly heterogeneous in their overall molecular composition and there are distinct molecular synapse profiles~\citep{Unterauer2024}. This molecular diversity likely extends to the lipid composition and mechanical state of the synaptic membrane, ultimately raising the question whether these different molecular synapse types also regulate stability and turnover differently.

Cell adhesion molecules are also important for synapse stability. For example, disrupting neurexin signalling increases synapse elimination and also other CAMs mediating trans-synaptic adhesion, such as cadherin, modulate synapse stability~\citep{Mendez2010, Duncan2021, Bian2015}. 

Apart from their role in organizing the intra-cellular structures, they also mediate forces between the membranes they are embedded in and are thereby regulating spine morphology in coordination with the intra-synaptic actin. This indicates a \emph{cell-mechanics-based regulation of structural plasticity}~\citep{Kilinc2018}, that extends beyond a static neuron geometry~\citep{Stepanyants2002,Chklovskii2004,Stepanyants2005} considered so far.
Taken together, CAMs are essential to synapse formation, maturation and stabilization~\citep{Suedhof2021}.

\paragraph{Astrocytes, the extracellular matrix and microglia}

In many cases synapses do not only feature pre- and postsynaptic sites but are further contacted or enwrapped by astrocytes as well as an extracellular matrix~\citep[ECM;][]{Chelini2018}, which are also implicated in structural plasticity: 
Both astrocytes~\citep{Stogsdill2017, Tan2021} and the ECM~\citep{Nikonenko2003b, Webb2007, Hama2004, Lilja2018} feature further CAMs  that can help with the assembly and stabilization of excitatory synapses. Furthermore, astrocytes secrete factors that promote synapse formation (\cite{Choi2026}, e.g., thrombospondins:~\cite{Christopherson2005}; glypticans: ~\cite{Allen2012}; or Hevin:~\cite{Kucukdereli2011}), often through influencing the ECM. Disrupting the ECM increases synaptic turnover~\citep{Kelly2015} and it has even been proposed that the perineuronal net -- a specialized and relatively stable form of the ECM -- may restrict where synapses can be formed and thereby contribute to very long-term memory~\citep{Tsien2013, Lev2024}. 

Microglia also contribute to structural synaptic remodeling and have been implicated particularly in synapse elimination through recognition of molecular signals that mark synapses for elimination~\citep{Peet2020}, including phosphatidylserine exposure~\citep{Kurematsu2022} and complement-dependent mechanisms~\citep{Hong2016, Stevens2007}.

Taken together, the probability of synapse formation stabilization and elimination is influenced by  the local distribution and state of glia cells and the ECM - adding a \emph{non-neuronal component to the spatial heterogeneity of structural plasticity.}  However, because both glial cells and ECM themselves can remodel in response to neuronal activity~\citep{Bernardinelli2014, Dityatev2007, Singh2026}, this local heterogeneity may emerge dynamically from reciprocal neuron-glia-ECM interactions, with potentially important consequences for learning and memory.

\subsubsection{Open problems}

\paragraph{How is structural plasticity controlled?}
First, many of the above described factors relate to neuronal activity. This raises the question to what extent neuronal activity regulates structural plasticity in support of activity homeostasis, learning, and memory. At the same time, the involvement of glial cells and intracellular organelles suggests that structural plasticity may also be shaped by metabolic and resource constraints. Moreover, there is still a lot of uncertainty in predicting where synapses will be formed, stabilized or eliminated, and there is turnover even when activity is blocked. Thus, the perceived randomness of synaptic rewiring may stem from a combination of limited predictability due to an incomplete set of predictors as well as actual biological randomness implying that structural plasticity may only partly be controlled by the neuronal network. 

Addressing this question will require identifying molecular and cellular mechanisms that govern rewiring and determining how they interact with neuronal activity, local resource constraints, and network dynamics in order to complete the set of predictors. This will require the integration of experimental approaches combining longitudinal connectomics with multiplex nanoscale imaging~\citep{Unterauer2024}, quantitative proteomics, and protein-turnover measurements~\citep{Fornasiero2018, Mohar2025} with computational and theoretical modeling that links molecular and cellular mechanisms to their consequences at the network level.
Along this line, understanding how these mechanisms and the network dynamics change during aging and neurodegeneration~\citep{Kluever2022} will provide an important additional perspective.

\paragraph{Synapse-type-dependent mechanisms}
Due to their accessibility to longitudinal imaging, much of the work on structural plasticity has focused on excitatory synapses on dendritic spines. Yet, by now it is also known that inhibitory synapses are remodelling~\citep{Flores2015, Flores2014, Villa2016} and their remodelling is often coordinated with the excitatory structural plasticity~\citep{Chen2012, Garbett2026}. As network function also strongly depends on inhibitory connections~\citep{Mongillo2018}, it is clear that we also need to know how the above factors co-modulate inhibitory and excitatory structural plasticity and whether there are more factors to take into account. As inhibitory neurons are heterogeneous and the inhibitory plasticity seems to depend on the sub-type~\citep{Chen2015}, this will be a major effort that also requires a broad collaboration within the experimental community. Collaboration with computational neuroscientists can also provide insight on potential mechanisms to be studied and for the design of focused experimental setups where the role of inhibitory plasticity is easier to isolate. 

Furthermore, also non-chemical synapses --- gap junctions --- undergo activity- and transport-dependent changes~\citep{o2019design} and also their self-organization and effect on network activity remain to be clarified.

\paragraph{What are the effects on the network level?}
We need to better understand, how structural plasticity at the level of individual neurons shapes network dynamics and information processing, and how network-level organization emerges from local plasticity rules. In particular, it remains to be clarified which functional advantages arise from sparse and selectively structured connectivity rather than all-to-all communication (for example through modularization, reduced population-wide correlations), plasticity operating on slower timescales, and whether the apparent  stochasticity in rewiring may itself have functional advantages (for example by enabling the sampling of alternative connectivity configurations). Moreover, the network-level consequences of the spatial heterogeneity in structural plasticity described above, and the extent to which this heterogeneity is shaped by neural activity remain to be elucidated. Addressing these questions will require the integration of  experiments with computational and theoretical modeling work to bridge  scales and determine how local rules of structural plasticity give rise to emergent network organization, dynamics, and function (see Section~\ref{sec:function} for existing efforts).

\paragraph{Bridging between molecular and synaptic level}
A similar problem arises between the molecular level and the synapse level. Although we know that interfering with certain proteins, reactions or gene translation influences the architecture of the synapse and their internal organization, we need a better understanding of the pathways which give rise to this, how it impacts synapse formation, stabilization and removal, and whether we have discovered all relevant factors. To this end, the data from different experimental methods and different timescales can be integrated into generative models whose predictions can then be compared with the synapse dynamics from imaging at the synapse scale \citep{Lee2024, BonillaQuintana2020,Thomas2025,Negri2025, Safari2025}. Such models, in turn, may require knowledge of binding and polymerization energies, ion coordination and permeation, and protein interfaces or on how protein function is modulated through membrane tension or lipid composition. To obtain such knowledge, modeling on the molecular scale can be employed for predicting structure and dynamic at quantum-mechanical or more coarse-grained resolution~\citep{Burke2023, Heiser2025, Schackert2023, Albani2024, DiMarino2023,zheng2024modeling}. 
Thus also here, multiple scales need to be bridged, which will require a tight feedback between modelers and experimentalists.

\paragraph{Therapeutic use}
Finally, much of our mechanistic knowledge of structural plasticity stems from animal models or cultured neurons and frequently relies on experimental perturbations that cannot be directly translated into therapeutic interventions. For clinical translation, we therefore need to determine to what extent these mechanisms also operate in human neurons and neural circuits~\citep[cf.][]{Schuenemann2025,Lenz2024} and how they can be engaged by clinically applicable interventions. This includes pharmacological approaches as well as non-invasive neuromodulation, such as transcranial magnetic stimulation (TMS), for which intervention parameters can be systematically varied and linked to neural effects across cellular and network scales~\citep{Siebner2022,Massimini2026}. A major challenge will be to determine how such interventions engage structural plasticity and how molecular and cellular changes translate into network-level and clinically relevant outcomes. Addressing this translational gap will require close integration of studies in human tissue and experimental models with clinical research and computational approaches that bridge these different scales.

\section{What is the computational function of structural plasticity in biological networks?}
\label{sec:function}

\subsubsection{State of the art}

Measuring structural plasticity and its interaction with neural activity at the network level is highly challenging in experiments: either one can get a static picture of the full network~\citep[the so-called connectome, cf.][]{vandenHeuvel2016,Dorkenwald2024, Microns2025,Helmstaedter2026} or a dynamic (longitudinal) picture that is restricted to smaller sets of neurons and connections~\citep[cf.][]{Trachtenberg2002,Verstraelen2020}. 
Therefore, computational models are needed to unravel the complex interaction between neural activity and structural plasticity and bridge scales from  molecular and microscopic synaptic processes to cellular, circuit, and network dynamics.
The models may serve two major goals: 1.~bridging between data obtained with different methodologies and across different scales to test whether mechanisms identified at one scale can explain observations at another (e.g., connectivity statistics);
and 2.~extracting mathematical principles that generate testable theoretical models of network organization, dynamics, and neural coding.
Such principles are needed to fully understand the behavioral impact of structural plasticity, and are thereby also essential for clinical and technological translations (cf. the next section). 
The behavioral impact most prominently includes \emph{support for associative memory} that may be implemented by structural plasticity on the network and intra-cellular levels, which may be critical to enable the remarkable lifelong learning capabilities of animals~\citep{Parisi2019}.  
Related to this, the dynamic properties of structural plasticity can also be linked to the phenomena of representational drift and the organization of connectivity onto different dendritic branches.
In the following, we will discuss these different aspects in more detail.

\paragraph{Generating biologically relevant connectivities}
A good example for the first goal of computational models mentioned above (bridging between different methodologies) is the self-organization of biologically relevant connectivity. Static connectomes provide detailed statistics on structural connectivity, while longitudinal experiments reveal the principles governing how individual neurons adapt their connections, which can be incorporated into network models. It can then be tested whether these models reproduce the measured connectivity statistics.

Along this line, it has been found in experiments that there can be multiple synapses from one neuron to another and that this number of synapses follows a highly non-random distribution~\citep{Fares2009,Dorkenwald2022}. And while computational studies often use approximations considering either none or a single synapse from any one neuron to another, a number of models involving Hebbian or weight-dependent structural plasticity mechanisms have successfully explained these distributions~\citep{Deger2012,Fauth2015,Fauth2015b}.

Moreover, biologically realistic connectivity \emph{in vitro} and in rodent cortex features an overrepresentation of bidirectional connections and connected cell clusters within the same local network \citep{Song2005, Perin2011, Okujeni2017}. How such structures can arise from simple rules has also been targeted by several modeling studies. \citet{Miner2016} could show that activity-dependent functional and weight-dependent structural plasticity are able to produce clustered groups of strongly connected neurons and other connectivity motifs that are found in the cortex. \citet{DiazPier2016} could show that also homeostatic structural plasticity can reproduce aspects of cortical connectivity including coarse backbone structures of the connectivity among layers and neuron types.

Reproducing these statistics of connectivity is important as it in turn influences network dynamics and can make it more brain-like~\citep{Sporns2013,Hiratani2018,Hramov2021,Wein2021}.

\paragraph{Stable memory despite continuous turnover} Structural plasticity can produce long-lasting changes in network connectivity and may therefore contribute to the consolidation and maintenance of long-term memories. However, the transient nature of individual synaptic structures raises a fundamental question in neuroscience: How can the brain produce stable memories from inherently unstable synapses? To resolve this dilemma, theoretical frameworks propose concepts such as stable core connections, equifunctional pairwise connections (connections between specific neurons whose strengths remain stable, even though the physical synapses underlying the connections are constantly changing), and equifunctional networks (networks whose memory functions remain stable despite constant synaptic turnover, relying on strong connectivity within groups of cells or assemblies).

First of all, it is possible that a subset of stable spines form a stable core of connections, which could explain memory functionality in some regions of the cortex. However, this cannot account for all aspects of memory representations~\citep{Chambers2017, Mongillo2017,Acker2019}.

Equifunctional pairwise connectivity on the synaptic level is established by a dynamic equilibrium between synapse removal and the selective stabilization of suitable synapses through Hebbian structural plasticity~\citep{Deger2012, Fauth2015}. The resulting positive feedback leads to collective dynamics, where multiple synapses stabilize each other and enable the recruitment of even more connections. These dynamics not only preserve memory traces despite underlying structural turnover but also allow the network to achieve fast learning while ensuring slow forgetting~\citep{Fauth2015b}. 

Equifunctional networks extend this principle from individual neuron pairs to the network level. Here, the specific identity of connections or even of the presynaptic neuron is typically not critical for memory functionality. Instead, stability relies on the strength of synaptic connectivity within the group of relevant cells. For example, in feed-forward networks, Hebbian structural plasticity can stabilize new connections that match and support the receptive field of a postsynaptic neuron, and thereby preserves its response~\citep{Acker2019}.
In recurrent networks, cells can form a so-called cell assembly or ensemble, in which it does not matter exactly which connections there are, as long as activity can spread and persist sufficiently within the assembly \citep{Hebb1949,Palm2014,Holtmaat2016}. Here, Hebbian structural plasticity has been demonstrated to be an important mechanism for the formation and stabilization of such assemblies~(\cite{Fauth2019}, compare also \cite{KalleKossio2021,Manz2023}). 
Simplified forms of Hebbian structural plasticity, in which weak synapses get removed but new ones are added randomly, have been used for connectivity refinement, for example, to improve the performance of reward-modulated self-organization~\citep{Yuan2023}. 
Thus, also on the network level, the presence of some form of Hebbian structural plasticity can stabilize memory traces despite synaptic turnover.

Homeostatic structural plasticity has been considered in earlier studies mostly with respect to its balancing effect on neural activity \citep{Fauth2016,Tiddia2024}. Interestingly, however, \citet{Gallinaro2018} could show that cell assembly formation can also be achieved with a structural plasticity rule based on firing rate homeostasis~\citep[cf.][]{Butz2013}.
Stimulating a sub-group of neurons in a balanced recurrent network with structural plasticity between excitatory neurons first causes a small decay in synapse numbers, which is followed by a strong increase after stimulation that only decays slowly.
Thus, in that model, the random combination of pre- and postsynaptic elements implicitly evaluates correlations between neuronal sub-groups in a network.
This mechanism has more recently also been shown to account for the actual formation and recall of memory representations~\citep{Gallinaro2022}.
Also these assemblies exhibit much faster learning than forgetting, which is remarkable as they lack the above described self-stabilization mechanism emerging from Hebbian plasticity.

After the initial memory formation, replay mechanisms, that take place, for example, during slow-wave sleep, are known to be essential for memory consolidation in the brain~\citep{Rasch2013,Mizuseki2017,Dudai2015,Acker2019}. By repeatedly reactivating patterns of neuronal activity after learning, replay can strongly influence the impact of synaptic and structural plasticity. Especially for synapse formation, which is typically slow, the coordinated reactivation of such patterns may promote the strengthening and stabilization of memory-related synaptic connections~\citep{Yang2014}. On the other hand, unrelated synapses may further be weakened or eliminated during replay or during phases of other network activity (most importantly, during REM sleep;~\citealp{Dudai2015,Li2017,Mizuseki2017}). Thus, repeated replay of activity pattern on the network level may progressively stabilize and sharpen memory-related connectivity on the microscopic scale. Along this line, models of weight-dependent structural plasticity could indeed bridge between these scales as they give rise to self-organized replay during sleep and exhibit the respective memory strengthening~\citep{Fauth2019}. 

In summary, replay mechanisms are relevant for preserving and consolidating important structural changes while allowing ongoing remodeling of less important structures, thereby balancing memory retention with network flexibility. 
However, their link to the molecular mechanisms of structural plasticity, as well as their interaction with mechanisms of synaptic memory consolidation, which also occur at the network level \citep{Luboeinski2021,Okuda2020}, remain to be studied. % via integrated models.

\paragraph{Accounting for memory phenomena at the cognitive level} Multiple theoretical studies have tested the impact of structural plasticity on associative memory at the cognitive level. For instance, using simple auto- and hetero-associative neural networks,~\citet{Knoblauch2014} found that structural plasticity can reorganize synapses so that they serve particular memories. This substantially increases memory capacity (i.e., the number of patterns that can be stored in the network) as opposed to plasticity of synaptic weights, especially, in sparsely connected networks. Similarly, the model can also reproduce several cognitive phenomena: graded amnesia after brain damage due to stronger effective connectivity of recently consolidated memories, resistance to catastrophic forgetting due to learning that ends just before the critical capacity limit is reached, and the spacing effect because distributed rehearsal allows new synapses to be grown and used for the learning. Similarly, in the feed-forward network model proposed by~\citet{Tiddia2024},
rewiring enables increased memory capacity. They demonstrate that potentiation processes can help to prevent connections from being pruned, which reduces catastrophic forgetting in a similar way as described by \citet{Knoblauch2014}. 

\paragraph{Link to representational drift} Representational drift refers to the finding that the neural substrate of memory representations changes (possibly randomly) over time while the memory content and retrievability remain largely unchanged~\citep{Rule2019,Micou2023}.
Understanding the relationship between structural plasticity and associative memory is linked to understanding representational drift, as it raises a similar question: How can memory or behavior remain stable even when neural activity patterns change significantly over time? 

While there is up to now no explicit model considering both structural plasticity and representational drift simultaneously, some existing models can yield estimates of the expected behavior:
On the one hand, the drifting assemblies model by Memmesheimer and others~\citep{KalleKossio2021,Manz2023} demonstrates that Hebbian synaptic plasticity dynamics can act to induce and compensate drift by driving weights to essentially zero or non-zero states, which resembles in effect the presence or absence of synapses in structural plasticity. Complementarily, another recent model~\citep{Eppler2026} attributes drift to the interaction between Hebbian learning and activity-independent synaptic stochasticity -- also much akin to the effects of structural plasticity. 
These models outline ways how drifting representations could emerge, but would have to be extended to explicitly account for structural plasticity. 
On the other hand, in the weight-dependent structural plasticity model by~\citet{Kappel2018}, spontaneous drift emerges from the exploring and sampling of degenerate network configurations that maximize the reward conveyed by a global signal. Thereby, this model describes equifunctional network states similar to those discussed above.

Regarding the functional impact of drifting representations,~\citet{Acker2019} could show that even without reward or error feedback, self-organized synaptic plasticity of the readout connections can mitigate the functional impact of random rewiring, and thus likely also of drifting representations.

Taken together, structural remodeling may, on the single synapse level, contribute to the emergence of representational drift, while also mitigating its consequences on the network-level. Yet, this relationship remains to be tested in models and experiments more explicitly. 

\paragraph{Organization of intra-neuronal connectivity to improve memory}
\label{sec:dendrites}
Dendritic spikes can provide powerful local signals for the induction of synaptic plasticity~\citep{Kampa2007}. \citet{Larkum1999} originally demonstrated that spatially and temporally coordinated synaptic inputs can generate dendritic spikes, providing powerful local signals to selectively amplify present activity. 
Based on this, \citet{Poirazi2001} showed that active dendritic branches can function as thresholded sub-units where synapses clustered on the same branch interact in a nonlinear manner, allowing each branch to respond to distinct input combinations and thereby greatly increasing the number of patterns that the neuron can detect.
This is supported by the finding that weight-dependent structural plasticity can cluster synapses responding to the same input on individual dendritic branches, which increases pattern recognition capacity~\citep{Limbacher2020}.
A possible mechanism behind this is provided by the calcium-dependent structural plasticity model from~\citet{Helias2008}, which assumes that calcium concentration in dendritic spines can act as a correlation detector between pre- and postsynaptic spikes~\citep[cf.][]{Sjoestroem2001} that can directly influence synapse stability. In that way, correlations between inputs stabilize the respective connections through cooperation, which could take place on each dendrite individually.

In summary, intra-neuronal connectivity, i.e., the arrangement and interaction of synapses across dendritic structures, adds another dimension to memory storage and computation. Structural plasticity may exploit this subcellular organization by clustering functionally related inputs on dendritic branches, thereby increasing pattern-recognition and memory capacity. Similar principles are recently also being explored for machine-learning applications \citep{Chavlis2025}.

\subsubsection{Open problems}

\paragraph{New experimental approaches} While the above discussed modeling works already provide a link between local structural plasticity and network self-organization, their predictions on network dynamics need to be tested more explicitly. To this end, longitudinal, high-resolution approaches should be developed to monitor structural remodeling in large circuits during learning paradigms as well as across development and aging. This shift from static connectomes to dynamic connectivity trajectories is essential for understanding how neural circuits form, stabilize and reorganize over time. To achieve this, existing techniques such as longitudinal two-photon imaging, electron microscopy, and in vivo calcium imaging could be extended to enable tracking the synaptic connections within a larger neural circuit over weeks or months. Alternatively, massive calcium imaging could be used to reconstruct both activity and structural connectivity from the same measurements \citep[see, for instance,][]{Stetter2012}. Furthermore, recently identified factors that are relevant for structural plasticity, such as astrocytes and microglia, might be imaged in parallel.  The obtained data should reveal how structural changes correlate with behavioral memory features as well as their consolidation and decay. Integrating such longitudinal data with complementary high-resolution datasets across spatial and temporal scales, will enable direct testing and refinement of  theoretical models, thereby helping to identify principles that link structural remodeling to neuronal activity and function.

\paragraph{Interaction with other types of plasticity} For a fundamental understanding of the brain, it is crucial to study how structural plasticity interacts with other types of plasticity, especially, because plasticity processes can share certain signaling pathways or exert concurrent changes.
Late-phase synaptic plasticity (referring to the late phases of LTP and LTD) is of particular importance here, as its timescales overlap with the timescales of structural plasticity and as it depends on morphological changes of the synapse~\citep{Redondo2011,BonillaQuintana2021b,Negri2025,Thomas2025}.

To study the complex interaction of late-phase synaptic plasticity and changes of the topology at the network level, integrated models need to be devised. To this end, future studies can combine existing network models for both types of mechanisms \citep[e.g.][]{lu2019network,Luboeinski2021,Luboeinski2022,Gallinaro2022}. For similar reasons, the interaction between structural plasticity and synaptic scaling mechanisms will also require approaches with integrated models~\citep[cf.][]{lu2025interplay}.

Complementarily, models on the molecular and subcellular scales should investigate the molecular signaling cascades that link synaptic activity to structural outcomes (e.g., actin remodeling and protein synthesis; \cite{Smolen2019,BonillaQuintana2021b,Negri2025,Thomas2025}). Such models could generate experimentally testable predictions about the co-regulation of synaptic rewiring and changes in synaptic strength, which can, in turn, inform models at the network scale. Integrating these scales should ultimately help to explain cognitive phenomena, such as memory consolidation, interference, and reconsolidation.

\paragraph{The emergence and impact of dendritic clustering} As discussed in Section~\ref{sec:regulation}, the emergence of synapse clustering on dendrites \citep{Fu2012,Chen2012} poses the fundamental question whether it is driven by the local activity in the dendrite or by molecular cues, or even by a combination of both. To elucidate this, multiscale computational models are needed that link local molecular and dendritic processes to network activity and the self-organization of synapses on dendritic branches. 
Existing network models incorporating dendritic compartments with non-linear integration \citep[see, e.g.,][]{Brunel2004,Kastellakis2016,Luboeinski2026} could be extended by incorporation of molecular cue-driven mechanisms. 
This may help to identify the activity patterns and molecular mobility regimes under which dendritic clustering occurs. Moreover such models could  reveal how dendritic clustering enhances computational efficiency in neural circuits \citep[cf.][]{Larkum1999,Poirazi2001,Poirazi2003,Helias2008,Limbacher2020}, which may in turn be validated in new experiments.

\paragraph{The influence of rewiring on network and whole brain activity}
A further open question is how structural plasticity at the single neuron scale influences activity and function on the scale of large networks or even whole brains. To incorporate structural plasticity mechanisms into large-scale computational models at feasible computational cost, molecular, synaptic, and cellular processes need to be abstracted into effective, biologically grounded rules that preserve those features that are critical for network-level dynamics.
Such rules should capture key interactions identified above, including Hebbian and homeostatic mechanisms of structural plasticity and their coupling to changes in synaptic efficacy.

Embedding such effective rules in multi-scale simulation frameworks may provide a route from local mechanisms of structural plasticity to circuit and ultimately whole-brain dynamics, enabling model predictions to be compared with data obtained in clinical settings~\citep[cf.][]{Hater2025}.
For clinical translation, it will further be important to determine how structural plasticity is modified by therapeutic interventions and how these changes affect the reciprocal interaction between connectivity and activity across scales. Addressing this challenge will require close interaction between experimental neuroscientists, computational modelers, and clinicians.

\section{What can we gain from understanding structural plasticity?}
\label{sec:gain}

\subsection{Clinical translation}
\label{sec:clinical_translation}

Alterations in neuronal structure and structural plasticity have been reported across a range of neurodevelopmental and neuropsychiatric disorders, including Autism Spectrum Disorder (ASD, \cite{hutsler2010increased}), epilepsy~\citep{rossini2021dendritic,multani1994neocortical,jiang1998spine}, and schizophrenia~\citep{moyer2015dendritic, macdonald2017selective}. Converging evidence further indicates that disease-associated genetic and molecular mechanisms can affect pathways involved in structural plasticity, suggesting that these mechanisms may represent potential targets for therapeutic interventions~\citep{forrest2018dendritic}. A major translational challenge is therefore to determine whether and how structural plasticity can be therapeutically engaged to promote adaptive circuit remodeling and functional recovery. This will require linking intervention-induced changes at molecular and cellular scales to their consequences for circuit and network function and, ultimately, clinical outcomes.

\subsubsection{State of the art}
\paragraph{Homeostatic structural plasticity and predictions for stroke therapy} The so-called model of structural plasticity (MSP), later on referred to as homeostatic structural plasticity (HSP), provided an early computational framework for investigating how activity-dependent structural remodeling may contribute to network reorganization after lesions~\citep{butz2009model,Butz2009,Butz2013,butz2014homeostatic} and has been adapted for large scale simulations~\citep{DiazPier2016}. In these models, neurons regulate the formation and removal of synaptic elements according to deviations of neuronal activity, represented by intracellular calcium, from a homeostatic target.

Following deafferentation, reduced activity in the lesion projection zone drives structural remodeling according to these homeostatic rules. Depending on the parameters governing axonal and dendritic growth, the model predicts sequential reorganization of connectivity and recovery of activity, including ingrowth of connections from neighboring intact regions \citep{butz2014homeostatic}.

Another prediction of the model is that adult networks have problems with compensating stroke-induced deafferentiation and that stimulation can help to mitigate this shortcoming~\citep{butz2009model}. Along this line, the model was used to test which stimulation patterns are more effective than others.

These studies illustrate how computational models can link local rules of structural plasticity to large-scale network reorganization after brain injury and generate hypotheses about how therapeutic interventions, including brain stimulation, might influence adaptive rewiring.

\paragraph{Brain stimulation}
Brain stimulation provides a unique experimental and translational approach to structural plasticity because stimulation parameters can be systematically controlled while plasticity-related changes can be examined across molecular, synaptic, cellular, and network scales. Animal models using transcranial direct current stimulation (tDCS), repetitive transcranial magnetic stimulation (rTMS), or deep brain stimulation (DBS) have demonstrated structural changes, including changes in spine density and spine turnover dynamics. 

For TMS, the immediate neural effects depend on the interaction between the induced electric field, neuronal morphology and excitability, and the functional state of the targeted network~\citep{Siebner2022, Massimini2026}, while repetitive stimulation can subsequently engage longer-lasting synaptic and structural plasticity mechanisms. Using mouse \textit{in vivo} studies,~\citet{barbati2020enhancing} demonstrated that repetitive direct current stimulation in mice increased the frequency of miniature excitatory postsynaptic currents (mEPSCs) and was later confirmed as increased spine density. A similar tendency was observed in magnetic stimulation but using organotypic tissue cultures of the mouse hippocampus. Both amplitude and frequency of mEPSCs increased and decayed during the 8-hour time window after a single session of stimulation~\citep{vlachos2012repetitive}. Along that line, \citet{lu2025repetitive} showed that although frequency changes in mEPSCs indicate changes in synapse number, the actual spine density probed by time-lapse imaging experiments did not show any increase. This opens up new discussions about whether imaging techniques capture structural plasticity or a hybrid view of structural plasticity should be formed by integrating evidence from both electrophysiological recordings and optical imaging experiments.

Computational modeling of brain stimulation opens up an alternative approach to disentangle structure and function: testing whether structural plasticity rules explain more empirical evidence than functional plasticity rules. One example from~\citet{lu2019network} used an excitation-inhibition balanced network and a linear homeostatic structural plasticity rule. It exhibited intriguing network remodeling phenomena induced by weak direct current stimulation. The model qualitatively predicts the impact of tDCS parameters (e.g., stimulation intensity, current polarity, stimulation focality, and repetitive patterns), and  that strong and focal stimulation triggers prominent network remodeling. In particular, instead of delivering an extensively long epoch of tDCS stimulation, partitioning the total stimulation duration with small intervals works more efficiently in triggering network remodeling. 

One natural follow-up is the modeling of transcranial magnetic stimulation (TMS), which is considered more focal than tDCS due to the magnetic field distribution of a typically used figure-eight coil~\citep{mosayebi2021impact}. 
\citet{anil2023repetitive} modeled TMS as a short square-shaped electric pulse at a single neuron and demonstrated that the higher intensity of TMS stimulation at supra-threshold (able to trigger a silent neuron to fire action potentials) does not guarantee stronger after-effects in terms of network remodeling, while  strong stimulation within the intensity range of tDCS (sub-threshold level, i.e., unable to trigger action potentials in silent neurons) achieves stronger network remodeling. \citet{lu2025resolving} further suggested that the effects vary if the same tDCS protocol is delivered to a brain region under different conditions -- that is, at resting state or in the process or aftermath of learning or memory consolidation -- thereby resolving the inconsistent experimental observations that tDCS administered at different stages of motor learning can both boost or hinder learning effects.

\subsubsection{Open problems}

\paragraph{Disease and set-point} Given the promising application of the model of homeostatic structural plasticity in elucidating spine dynamics in disease and in therapeutic approaches, it is of great importance to pinpoint the exact rule of structural plasticity in disease conditions. Experimental studies have demonstrated firing-rate homeostasis following prolonged sensory deprivation, with neuronal activity returning towards pre-perturbation levels over time  \citep{hengen2013firing,hengen2016neuronal}. Importantly, the dynamics of this recovery depend on neuronal cell type and brain state, and firing-rate homeostasis can persist under pathological conditions, although other aspects of network dynamics may be altered \citep{mcgregor2024failure}. These observations provide important experimental constraints for models of homeostatic structural plasticity, but whether the set points governing structural remodeling remain unchanged in disease is largely unknown. Recovery of firing rates after perturbation provides an important constraint for such models, but does not by itself establish that the same set point directly governs structural remodeling. Determining this relationship seems to be one of the key open questions.

A major open question is therefore whether homeostatic target activity is defined at the level of individual neurons or emerges at the population or network level, and whether these target values remain stable or adapt following injury or disease. Experimental observations from different brain regions and species provide evidence for different forms of activity homeostasis ~\citep{hengen2016neuronal,slomowitz2015interplay}, emphasizing the need to determine which homeostatic variables and target values are relevant for structural remodeling in specific biological contexts.

\paragraph{Is structural plasticity sufficient to explain the whole story?}
Despite the explanatory power of homeostatic structural plasticity in models of brain stimulation and post-lesional reorganization, structural remodeling does not operate in isolation. Also here, the question is how structural plasticity interacts with other forms of plasticity operating at different timescales, and how these interactions shape network stability and adaptive remodeling. For example ~\citet{manos2021long} combined homeostatic structural plasticity with the fast dynamics of STDP  and showed that their interaction can facilitate network reorganization and explain the prominent effects of coordinated reset stimulation (CRS), which was originally used in deep brain stimulation \citep{tass2003model} and now introduced to vibrotactile stimulation for treating Parkinson's disease \citep{tass2017vibrotactile}. 

Similarly, models combining homeostatic structural plasticity with homeostatic synaptic scaling suggest that their interaction depends on network state: the two mechanisms can be partly redundant under moderate perturbations but become complementary under strong activity deprivation \citep{lu2025interplay}.
A complementary model study demonstrates how firing-rate homeostasis counteracts synapse loss in Alzheimer's disease~\citep{Bachmann2020}; they show that the network’s dynamical characteristics and sensitivity can be recovered by strengthening the remaining synapses.
These findings illustrate that structural remodeling cannot necessarily be understood in isolation, but may emerge from interactions between plasticity mechanisms operating at different timescales. An important challenge will therefore be to determine how structural, synaptic, and other forms of plasticity interact across timescales to balance network stability with adaptive remodeling, and how these interactions are altered in disease or by therapeutic interventions.

\subsection{Large-scale simulation and neuromorphic applications}
\label{sec:simulations}

Simulating a ``digital twin" of the human brain~\citep{Amunts2024} at a level of abstraction that is useful for a given problem statement is a goal shared by neuroscientists for understanding brain function through the study of virtual brains, clinicians for designing therapies, and neuromorphic engineers for developing energy-efficient devices.
Since structural plasticity appears to be ubiquitous in the brain and links molecular processes with large-scale network structure across temporal scales, the detailed processes need to be understood in order to find the right levels of abstraction for specific applications.
Developing scalable algorithms for high-performance computing systems that enable large-scale simulations for brain research or next-generation, neuro-inspired AI may contrast the simulation technology required for low-power, memory-restricted neuromorphic chips and edge devices.
However, both ends of the spectrum benefit from a proper understanding and application of structural plasticity, for example, to devise efficient algorithms that capture the statistics of dynamically changing connectivity, or to enable running machine-learning models on systems with a limited number of synapses available.

\subsubsection{State of the art}
\paragraph{Simulation technology for neuroscience}

For a translation to spiking neural networks, simulation technology for neuroscience needs to support the functionality to express structural plasticity rules and to run long simulations efficiently.
Especially large and complex network models have requirements in terms of reproducibility, extensibility, and efficiency that cannot be solely fulfilled by general-purpose programming languages such as MATLAB, Python, or C/C++~\citep{Senk2022,Plesser2025}, and therefore rely on dedicated neural network simulators.
Nowadays, several simulators exist that differ not only in their targeted software and hardware platforms (e.g., conventional vs. neuromorphic technologies), but also in their main level of description of the biological system (e.g., large networks of point-neurons vs. morphologically detailed neurons), and they come with different design choices and different constraints.

Running a simulation typically includes a network construction phase in which nodes and connections are created in computer memory, followed by a state propagation phase~\citep[illustrated for the simulator NEST by][Fig. 3]{Albers2022}.
Some simulators provide high-level routines for defining network graphs based on connectivity concepts derived from biological networks.
These routines allow for the generation of dense or sparse connectivity patterns between neuronal populations according to deterministic or probabilistic rules~\citep{Senk2022,KorcsakGorzo2024}.
In the state propagation phase, the connectivity is typically fixed during updates of the states of neurons and synapses (e.g., the synaptic weight in case of synaptic plasticity).
Manipulating existing connections during the state propagation phase, as required for structural plasticity processes, is challenging and computationally costly because the data structures for storing synapses need to be touched during runtime.

A number of simulator developers have already tackled this challenge and demonstrated simulations of structural plasticity:
Starting with conventional CPU-based technologies, the simulator NEST currently provides the functionality to add and remove synapses outside of the state propagation phase, and a fully integrated implementation of the homeostatic structural plasticity rule by~\citet{Butz2013} that runs without interrupting the simulation kernel~\citep{DiazPier2016}.
A scalable approximation algorithm for that rule has been proposed for large-scale simulations~\citep{Rinke2018,Czappa2023, Czappa2026}.
Rewiring is also supported by the simulator ANNarchy for its CPU backend; here synapses can be added or removed based on a condition~\citep{Vitay2015}.
The simulator Arbor provides a looser integration in a backend agnostic approach~\citep{arbor-structural-plasticity}, while the simulator Brian~2 can only adjust a synapse status during runtime without actually adding or removing synapses~\citep{brian-structural-plasticity}.
For GPU-based systems, a flexible framework for building structural plasticity rules has recently been added to GeNN. It offers a flexible interface for defining structural plasticity rules and demonstrated efficient simulations in spike-based machine learning and computational neuroscience contexts ~\citep{Knight2026}.

On neuromorphic systems, one example combines analog VLSI with an FPGA-based structural-plasticity co-processor~\citep{George2017}.
Similarly, a custom embedded digital processor located on chip handles rewiring for the analog neuromorphic system BrainScaleS-2~\citep{Billaudelle2021}.
The digital neuromorphic hardware SpiNNaker shows structural plasticity both with SpiNNaker1~\citep{Bogdan2018} and SpiNNaker2~\citep{Liu2018,Yan2019}.

To conclude, simulator developers established different approaches for implementing structural plasticity, ranging from updating the status of existing synapses to dynamically adding and removing connections, and implementing biologically motivated structural plasticity rules. A major challenge is now to determine which biological features of structural plasticity need to be retained at different levels of abstraction to reproduce its consequences for network organization and function, while remaining computationally tractable at large scales.

\paragraph{Machine learning with limited synapse numbers}
Machine learning models traditionally go through the consecutive phases of training first and inference later: an artificial neural network with dense (i.e., mainly all-to-all) connectivity learns from data by adjusting its weights, and only once this is completed, the final network is put to work.
Deep learning has high computational demands and usually relies on GPU-powered high-performance computing systems.
With the aim to make these models more efficient or even deploying them on systems with limited resources, novel algorithms are being explored.
For instance, neuromorphic hardware or edge devices often have limited memory available and therefore cannot store all synaptic weights~\citep{Billaudelle2021} of a large model.
To reduce the amount of memory required, one approach is to reduce the weight resolution (e.g.,~\cite{Gupta2015}) and another is to reduce the number of synapses.

One way to achieve this is pruning~\citep{Gale2019}, that is training a dense network and then removing weak synapses for inference without compromising on the accuracy.
To optimise resource allocation already during the training phase, structural-plasticity inspired learning rules can help to train networks that are already initialized with sparse connectivity~\citep{Billaudelle2021}.
Training sparse networks requires different algorithms, but can achieve similar accuracy as dense networks at significantly reduced training time~\citep{Knight2026}, for example, using the algorithm DEEP~R~\citep{Bellec2018} that replaces weak synapses with randomly sampled new ones.
Such abstract structural plasticity rules can also be combined with biologically plausible learning rules for updating the synaptic weights (e.g.,~\citet{Bellec2020,Wunderlich2021}).
Yet another approach is to grow so-called constructive neural networks during training~\citep{Frean1990}.

Taken together, these results demonstrate the potential of structural-plasticity-inspired principles for dynamically allocating limited connectivity and computational resources in artificial neural networks.

\subsubsection{Open problems}

\paragraph{Efficient large-scale simulation technology}

Since structural plasticity is overall a slow process, accelerated simulations are needed to validate models and complement experimental work.
It is already expensive to store large numbers of synapses with their parameters or state variables (e.g., the weights) in computer memory, and dynamically changing connectivity-related data structures during a simulation of structural plasticity is even more costly~\citep{Knight2026}.
To date there are also no common benchmarks or frameworks for structural plasticity comparing different simulation technologies~\citep{Yik2025,Senk2026}.
A co-design of structural plasticity models and simulation technology is required to maintain the generally desired separation of models from simulators~\citep{Einevoll2019}, while developing efficient algorithms for specific hardware- or software-based simulators.
To streamline these developments, conceiving overarching roadmaps for structural plasticity technology should be a joint task of different simulator developers together with theoretical and experimental neuroscientists as well as experts for clinical translation and neuromorphic applications. Along this line, as a community it is essential to build on the already existing approaches and jointly define the next design goals for tools to maximise the potential for tool sharing and interaction for common pipelines.

Bridging between disciplines will further require extending structural plasticity simulations beyond point-neuron networks towards models that incorporate relevant aspects of neuronal morphologies. Tools such as NEAT~\citep{Wybo2026} and the Trees Toolbox~\citep{Cuntz2010} provide starting points for incorporating detailed dendritic and axonal structures. As discussed above, a major challenge for this will be to determine which biological details need to be retained to capture the relevant consequences of structural plasticity while keeping simulations computationally tractable at large scales.

\paragraph{Training algorithms for sparse networks}

Sparsity in network connectivity brings advantages like memory efficiency and potentially reduced training time, but requires the development of novel training algorithms beyond classical machine learning methods.
Possible research directions include, for example, the incorporation of modulatory factors derived from top level objectives, such as neuromodulation, reward signals or energy constraints~\citep[cf.][]{Bellec2018}.
Based on the predictions from models and large-scale simulations using algorithms, the biological implementation of these top-level objectives can then be tested in experiments.
Similarly, structural-plasticity-inspired approaches may allow network connectivity to adapt during training rather than being predefined and fixed, potentially improving network function and resource efficiency~\citep{Achterberg2023}. 
According to the ``lottery ticket hypothesis"~\citep{Frankle2018}, dense networks contain sparse sub-networks (winning tickets) which can achieve the same test accuracy as the dense networks if trained in isolation in a similar number of iterations.
To construct a structural plasticity algorithm that finds these winning tickets it is crucial that experts from neuroscience, machine learning and neuromorphic hardware come together and develop efficient algorithms inspired by the brain's structural plasticity accounting for potential limitations of the hardware such as the availability of memory.

\addsec{Conclusion}

In the recent years, major advances have been made in understanding the regulation of structural plasticity at molecular, synaptic, and cellular scales, as well as its consequences for circuit and network organization. At the same time, structural plasticity has become increasingly relevant to questions ranging from long-term memory, representational drift, and dendritic computations to clinical interventions, large-scale brain simulation, and resource-efficient artificial and neuromorphic systems.

A central conclusion emerging from these developments is that structural plasticity is inherently a multiscale problem. Molecular and cellular mechanisms determine where and when structural remodeling can occur, while its functional consequences emerge at the level of neurons, circuits, and networks over timescales ranging from minutes to months or longer. Understanding structural plasticity therefore requires more than studying these individual scales in isolation: it requires determining how mechanisms identified at one scale give rise to organization and function at another.

Recent advances in longitudinal imaging, connectomics, molecular and physiological approaches, computational modeling, and simulation technology now provide complementary means to address this challenge. Experiments can identify and constrain biological mechanisms, computational models can bridge spatial and temporal scales and generate predictions about their emergent consequences, and experimental and translational applications can test these predictions and reveal which aspects of structural plasticity are functionally relevant.

We therefore propose that progress in understanding and exploiting structural plasticity will require a strongly integrated, iterative approach connecting experiments, computational models, and applications across spatial and temporal scales. Such a framework should enable us to identify general principles of adaptive structural remodeling, determine how they contribute to brain function in health and disease, and establish when these principles can be translated into therapeutic and technological applications.

\newpage
\addsec{Acknowledgements}

This manuscript is based on the ideas developed during the scoping workshop ``Structural Plasticity in Networks of the Brain: Experiments, Models, Applications" funded by the Volkswagen Foundation (VolkswagenStiftung). 

The work was supported by
the German Research Foundation (Deutsche Forschungsgemeinschaft, DFG) through grants SFB1286/C2 (to MF), SFB1286/C12 (to JS, EFF), SFB 936/A1/Z3 (to MP, CH),  SPP2041-313856816 (HI1286/7-1; to MP), TRR169-261402652 (A2; to MP), HU937/27-1 (to MP), CU 217/7-1 (to HC), NeuroCure EXC-2049 – 390688087 (to AW), FOR 2705 (to GT), RTG 2416 (to GT), and 545776403/FOR5880 (Mod4Comp; to SK);
the German Federal Institute for Risk Assessment through grant 60-0102-01.P636 (to HC);
the German Federal Ministry of Research, Technology and Space (BMFTR) under grant number 01EE2503E (to SH);
EPSRC Grants EP/V052241/1 and EP/S030964/1 (to JK);
the National Institute on Aging of the National Institutes of Health (NIH) under award number R21AG085062 (to EFF);
the Volkswagen Foundation (VolkswagenStiftung) through project number 9D558/-1 (to JS, SK); and
the Helmholtz Joint Lab ``Supercomputing and Modeling for the Human Brain" (to SD, HL, JL, VA, JS, AM).
The authors would like to thank Prof. Christoph von der Malsburg for fruitful discussions regarding the role of structural plasticity in the future of neuro-inspired AI.

\addsec{References}
\AtNextBibliography{\small}
\printbibliography[heading=none]

\end{document}